# Graphene Synthesis via the High Pressure – High Temperature Growth Process

F. Parvizi[1], D. Teweldebrhan[1], S. Ghosh[1], I. Calizo[1], A.A. Balandin[1,2,*], H. Zhu[3], R. Abbaschian[2,4]

[1]*Nano-Device Laboratory, Department of Electrical Engineering, University of California – Riverside, Riverside, California 92521 U.S.A.*

[2]*Materials Science and Engineering Program, Bourns College of Engineering, University of California – Riverside, Riverside, California 92521 U.S.A.*

[3]*The Gemesis Corporation, 7040 Professional Parkway East, Sarasota, Florida 34240 U.S.A.*

[4]*Department of Mechanical Engineering, University of California – Riverside, Riverside, California 92521 U.S.A.*

## Abstract

We report on a new method for graphene synthesis and assessment of the properties of the resulting large-area graphene layers. Graphene was produced by the high pressure – high temperature growth from the natural graphitic source by utilizing the molten *Fe-Ni* catalysts for dissolution of carbon. The resulting large-area graphene flakes were transferred to the silicon – silicon oxide substrates for the spectroscopic micro-Raman and scanning electron microscopy inspection. The analysis of the *G* peak, *D, T+D* and *2D* bands in the Raman spectra under the 488-nm laser excitation indicate that the high pressure – high temperature technique is capable of producing the high-quality large-area single-layer graphene with a low defect density. The proposed method may lead to a more reliable graphene synthesis and facilitate its purification and chemical doping.

---

[*] Corresponding author; electronic address (A.A. Balandin): balandin@ee.ucr.edu ; http://www.ndl.ee.ucr.edu/

F. Parvizi, D. Teweldebrhan, S. Ghosh, I. Calizo, A.A. Balandin, H. Zhu and R. Abbaschian, UCR 2008

The electronic properties of graphene, which is a carbon sheet of a single or few-atomic-layer thickness, attracted much attention owing to its unusual physical properties and the possibility of electronic applications [1-5]. It was suggested that a band gap can be engineered in a single-layer graphene by the spatial confinement [5]. Graphene's extremely high room-temperature carrier mobility makes it a promising material for applications in the future nanoelectronic circuits. A number of graphene – based devices have been proposed theoretically or tested [6-9]. It is expected that graphene can be more easily integrated with the standard Si complementary metal-oxide-semiconductor technology than carbon nanotubes owing to graphene's planar geometry. A recently discovered superior thermal conductivity of graphene [10], which exceeds that of carbon nanotubes and diamond, establishes graphene as a thermal management materials and improves graphene's prospective as electronic material.

To achieve further progress in graphene electronics one has to develop a technology for producing a large area continuous graphene sheets with a controlled number of layers and low defect density. One also has to envision a reliable method for graphene chemical doping. As of today, the highest quality graphene layers were produced by the mechanical exfoliation of graphene from the highly oriented pyrolytic graphite (HOPG) with the help of some adhesive like a scotch tape [1]. The process of peeling of graphene from HOPG pieces is not well-controlled, provides only small size graphene flakes and hardly can be scaled up. The reported alternative techniques for producing graphene layers include the radio-frequency plasma-enhanced chemical vapor deposition (PECVD) [11], high-vacuum graphene growth by Si sublimation from 4H-SiC (0001) surface [12], electrostatic deposition [13], as well as chemical methods [14]. At the same time, the quality of graphene layers produced by the alternative techniques, available today, is not as good as that of the mechanically exfoliated from HOPG as evidenced from the low-temperature transport studies and Raman inspection.

Here we report a new approach for graphene synthesis using the high-pressure high-temperature (HPHT) growth [15] and describe the physical properties of the resulting material. The samples were synthesized in the split sphere growth apparatus, which is conventionally used for diamond growth [16]. The detail description of the apparatus has been reported by Abbaschian *et al.* [17]. To produce easily exfoliated graphitic layers we loaded bulk natural graphite on top of the solvent, which was placed on a substrate of stabilized zirconia infiltrated with *CsCl* (see Figure 1 (a-b)). Figure 1 (a) shows a schematic of the chamber and illustrates the growth process. In order to grow a monocrystal material we used metal catalyst consisting of *Fe - Ni* alloy and placed a small seed diamond crystal. During the growth process, the graphite source material completely dissolves in the solvent and re-grows on the substrate as either single- or polycrystalline diamond or graphitic layers, depending on the growth conditions.

Regardless of the catalyst used, the most critical requirement for producing the desired carbon material is the control of the temperature $T$ and pressure $P$ through the entire process. By varying the growth conditions, we found a regime which leads to the formation of the loose graphitic layers and flakes of graphene rather than diamond. The selected $T$ and $P$ values were in the range 5 – 6 GPa and 1300 – 1700 $^o$C, which correspond to the region right below the diamond – graphite equilibrium line in the HPHT phase diagram [17]. Excessive growth rates lead to inclusions and morphological instabilities resulting in strain and high density of defects. For this reason, we kept our mass growth rates low, at about 4 mg/h during the initial stage (less

F. Parvizi, D. Teweldebrhan, S. Ghosh, I. Calizo, A.A. Balandin, H. Zhu and R. Abbaschian, UCR 2008

than 40 h), and gradually increased them to about 20 mg/h at the final stage. Such a regime corresponded to an approximately constant deposition rate per unit area (mm/h).

The resulting material, removed from the growth chamber, has a cylindrical shape with ~ 1 cm diameter and height (see Figure 1 (b)). Although most of the cylinder material is of graphitic nature, the bottom re-crystallized carbon layers (up to ~1 mm) are very loosely connected to each other and can be easily peeled off. We have transferred a number of layers to Si/SiO$_2$ substrates without the help of the adhesives but by simply rubbing the cylinders against the substrates. In order to confirm the number of layers and assess the quality of HPHT graphene we used micro-Raman spectroscopy, which has proven to be the most reliable tool for graphene characterization [18-19]. It was not possible to carry out micro-Raman characterization of the top graphene flakes directly on the cylinders, removed from the growth chamber, because of the large background signal from the underlying carbon material. The obtained graphene flakes are large in size with many exceeding ~10 µm and some exceeding 20 µm in one of the lateral dimensions. Figure 2 is a scanning electron microscopy (SEM) image of a graphene sample, which was taken to verify the material uniformity.

The micro-Raman spectroscopy has been carried out using the confocal Renishaw instrument [19]. The spatial resolution in our experiments has been limited only by the laser spot size (~1µm), which allowed us to selectively examine different regions of the few-micrometer sized graphene layers. The spectra were excited by the 488 nm visible laser. An optical microscope with 50x objective was used to collect the backscattered light. The spectra were recorded with the 1800 lines/mm grating. A special precaution was taken to avoid the local excitation laser heating of the samples by keeping the power on top of the samples below 4 mW.

Figure 3 shows a typical spectrum from the HPHT single layer graphene. The most notable features of the spectrum are *G* peak at ~1580 cm$^{-1}$, which corresponds to the $E_{2g}$ mode, and a relatively wide *2D* band around 2703 cm$^{-1}$ [18-19]. The *2D* band is an overtone of the disorder-induced *D* band, which is frequently observed in carbon materials at ~ 1350 – 1360 cm$^{-1}$. The *D* band corresponds to the in-plane $A_{1g}$ *(LA)* zone-edge mode, which is silent for the infinite layer dimensions but becomes Raman active for the small layers or layers with substantial number of defects through the relaxation of the phonon wave-vector selection rules [20]. The band at ~2445 cm$^{-1}$ was attributed to the *T+D$_2$* combination similar to the one observed in the carbon-implanted HOPG [21] and graphite crystal edge planes [22]. While the presence of *D* band in the spectra taken from a graphitic sample away from its edges indicates the structural disorder or other defects, it is not the case for *2D* band. The second-order phonon *2D* band is always present in graphene and other carbon materials owing to the double-resonance processes [23] involving two phonons. The arrow in Figure 2 indicates the expected spectral position of the *D* band. The absence of the disorder band in the spectra of HPHT graphene and graphene multi-layers confirms the quality of the grown material.

To understand the evolution of the material during the HPHT growth process and determine the number of graphene layers we analyzed the *G* peak region and the shape of *2D* band in the initial graphite and resulting HPHT material. Figure 4 shows a close up of the *G* peak region for the initial bulk graphite loaded to the growth apparatus (marked by squares), HPHT as-grown graphitic layers (marked by triangles) and single layer graphene separated from the top of the HPHT cylinders (marked by circles). The

F. Parvizi, D. Teweldebrhan, S. Ghosh, I. Calizo, A.A. Balandin, H. Zhu and R. Abbaschian, UCR 2008

*G* peak position in graphene is slightly up-shifted as compared to graphite, which is in line with Refs. [18-19]. It is interesting to note that the disorder-induced *D* band at 1359 cm$^{-1}$ is very strong in the initial low grade natural graphite source material but completely absent in as-grown HPHT graphitic layers and single layer graphene (see also Figure 3).

The absence of the *D* band indicates that HPHT graphene has higher crystalline order and lower defect concentration. The latter was attributed to the fact that in HPHT process carbon is completely dissolved with the help of the molten *Fe – Ni* catalysts and purified during its transport to a growth site and following re-crystallization. One should note here that the stage when carbon is dissolved is suitable for its chemical doping. From the further examination of the figure, one can see that the *G*-peak intensity decrease as one goes from the initial bulk graphite to as-grown HPHT material, and to the single layer graphene is expected due to the reduction in the number of the interacting carbon atoms.

Figure 5 presents a *2D* band region for the same three forms of the material. Note that while the single layer graphene signature at 2703 cm$^{-1}$ is a sharp single peak, it is a broader band consisting of two peaks for the initial graphite source and as-grown HPHT graphitic layers. The main *2D* peak for the initial bulk graphite is located at 2718 cm$^{-1}$, which exactly corresponds to the doubled frequency of the disorder *D* peak. The second, lower peak, in the initial graphite source and HPHT graphite layers is seen as a shoulder around 2707 cm$^{-1}$. This observation is in line with the earlier reports that *2D* peak in bulk graphite consists of the two components *2D$_1$* and *2D$_2$* [20]. The observed sharp peak in the *2D* band for the single layer graphene is in agreement the previously reported data for graphene obtained by the HOPG exfoliation [18-19]. The <u>*2D* band and *G* peak features confirm that the HPHT produced sample is a single layer</u> graphene.

In conclusion, we reported on a new method for the high-pressure high-temperature growth of the high-quality large-area graphene layers from the natural graphite. The proposed method may lead to a larger scale graphene synthesis and a possibility of the efficient purification and chemical doping of graphene during the melt phase of the growth. In the growth process from the natural graphite we made use of the molten *Fe-Ni* catalysts for dissolution of carbon. The evolution of the disorder *D* peak as one goes from the initial source carbon to the graphitic layers and then to graphene suggests that the density of the defects reduces and the quality of material improves. The proposed HPHT method may lead to a more reliable graphene synthesis, larger area flakes and facilitate graphene's chemical doping.

*Acknowledgements*

A.A.B. acknowledges support from the DARPA – SRC Focus Center Research Program (FCRP) - Center on Functional Engineered Nano Architectonics (FENA) and the DARPA – DMEA funded UCR – UCLA – UCSB Center for Nanoscience Innovations for Defense (CNID).

F. Parvizi, D. Teweldebrhan, S. Ghosh, I. Calizo, A.A. Balandin, H. Zhu and R. Abbaschian, UCR 2008

**Figure Captions**

Figure 1: (a) Schematic of the graphene growth "cylinder"; (b) illustration of the high pressure – high temperature growth in the "split sphere" apparatus; (c) image of the cylinder with the material after it was removed from the growth chamber. Note that the region in the center, near the seed, is uniform indicating crystalline layers while the material at the edges is rough amorphous carbon.

Figure 2: (a) Optical image of the HPHT grown graphene flakes. The gray – green color regions correspond to single layer graphene while yellowish regions are bulk graphitic pieces. (b) Scanning electron microscopy image of the edges of a large graphene flake with the dimensions of ~ 10 μm x 3 μm produced via the HPHT growth. The SEM inspection confirmed the uniformity of graphene layers.

Figure 3: Raman spectrum of the single layer graphene (a) and few-layer graphene (b) produced via the HPHT melting growth process with *Fe-Ni* catalysts. The shape and location of the *G* peak and *2D* band were used to count the layers. The absence of the disorder *D* band near 1350 cm$^{-1}$ attests to the high quality of the grown graphene.

Figure 4: Raman spectrum of the initial natural bulk graphite source, as-produced HPHT graphitic layers and single-layer graphene in the *G*-peak region. Note that the disorder-induced *D* band is very strong in the spectra of the initial bulk graphite and it is completely absent in the spectra of graphene. This suggests the material quality improves during the growth process via melting and re-crystallization.

Figure 5: Raman spectrum of the initial natural bulk graphite source, as-produced HPHT graphitic layers and single-layer graphene in the *2D*-band region. Note that the *2D*-band from the single-layer graphene is symmetric while those from the bulk graphite and as-produced HPHT manifest a shoulder.

F. Parvizi, D. Teweldebrhan, S. Ghosh, I. Calizo, A.A. Balandin, H. Zhu and R. Abbaschian, UCR 2008

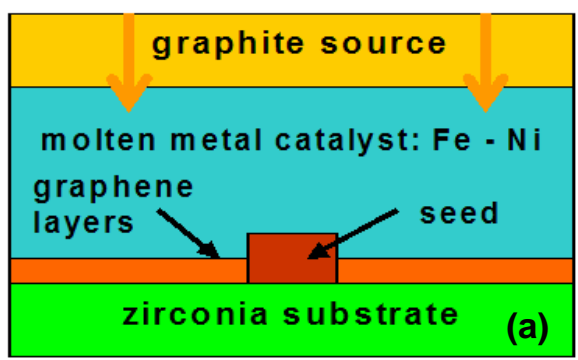

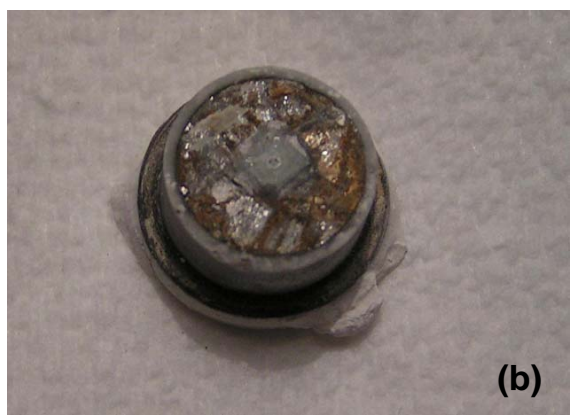

Figure 1 of 5

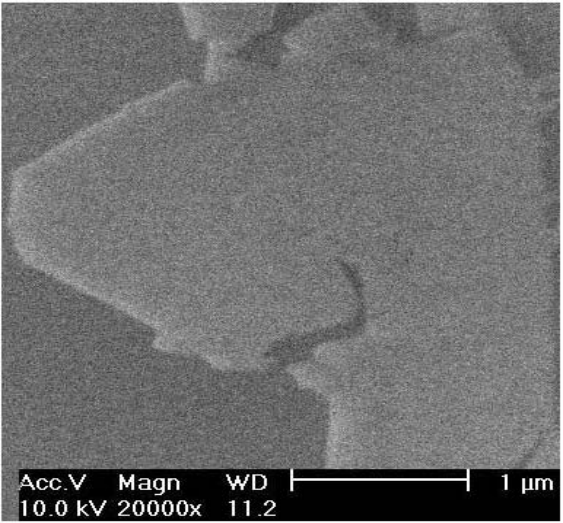

Figure 2 of 5

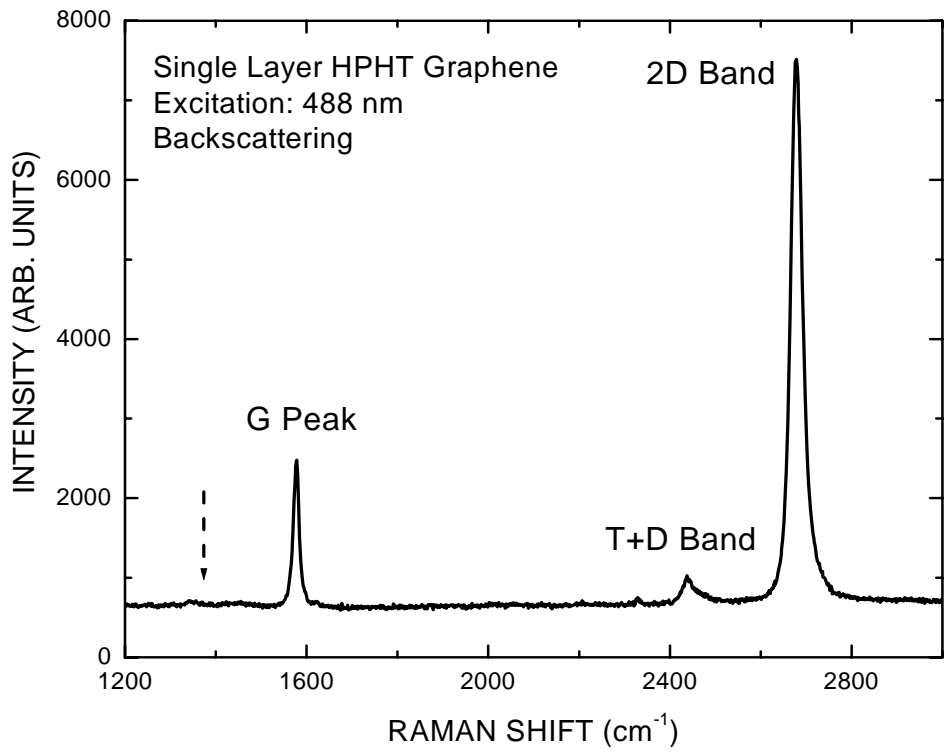

Figure 3 of 5

F. Parvizi, D. Teweldebrhan, S. Ghosh, I. Calizo, A.A. Balandin, H. Zhu and R. Abbaschian, UCR 2008

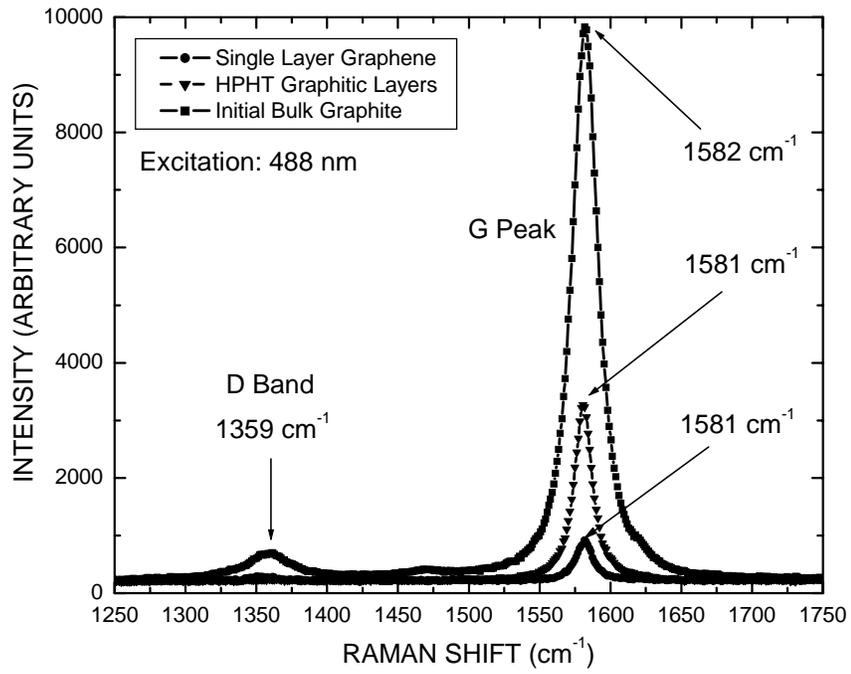

Figure 4 of 5

F. Parvizi, D. Teweldebrhan, S. Ghosh, I. Calizo, A.A. Balandin, H. Zhu and R. Abbaschian, UCR 2008

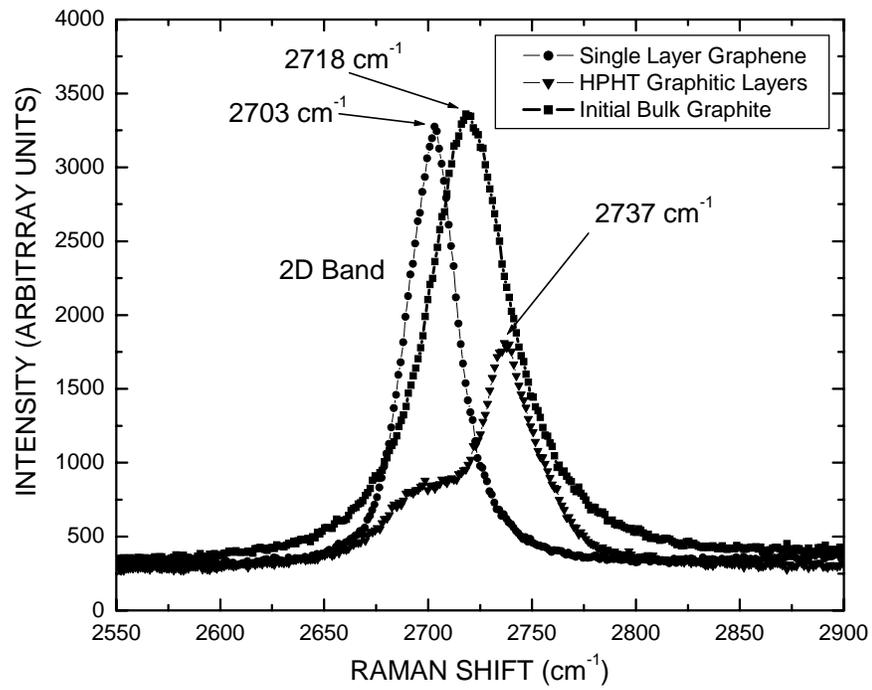

Figure 5 of 5